# STELLAR OCCULTATION OBSERVATIONS OF SATURN'S NORTH-POLAR TEMPERATURE STRUCTURE

A. R. Cooray[1], J. L. Elliot[2,3]
Department of Earth, Atmospheric, and Planetary Sciences
and
Department of Physics
Massachusetts Institute of Technology, Cambridge, MA 02139-4307

E-mail: asante@hyde.uchicago.edu, jle@mit.edu

A. S. Bosh[2,4]
Lowell Observatory
Flagstaff, AZ 86001-4499

E-mail: amanda@lowell.edu

L. A. Young
Center for Space Physics
Boston University, Boston, MA 02215

E-mail: layoung@bu.edu

M. A. Shure
Center for High Angular Resolution Astronomy
Georgia State University, Atlanta, GA 30303

E-mail: shure@chara.gsu.edu

[1]Now at Laboratory for Astrophysics and Space Research, University of Chicago, Chicago, IL 60637.

[2]Visiting Astronomer at the Infrared Telescope Facility which is operated by the University of Hawaii under contract to the National Aeronautics and Space Administration.

[3]Also at Lowell Observatory, Flagstaff, AZ 86001-4499.

[4]Now at Department of Physics, 133 Hofstra University, Hempstead, NY 11550-1090

ABSTRACT

We have observed a stellar occultation of GSC5249-01240 by Saturn's north polar region on November 20, 1995 from NASA's Infrared Telescope Facility (IRTF). This is the first recorded occultation by the polar region of a giant planet. The occulted region extends 88 km in vertical height and 660 km in horizontal length, over a region from 82.5 to 85 degrees in planetocentric latitude and from 20 to 30 degrees in planetocentric longitude. Based on isothermal model fits to the light curve, we find an equivalent isothermal temperature of $130 \pm 10$ K at a pressure level of $1.6 \pm 0.1$ µbar, which corresponds to a half-light latitude of $83.2 \pm 0.2$ degrees and longitude of $24.1 \pm 0.5$ degrees. Using numerical inversion procedures, we have retrieved the temperature profile of the occulted region, which suggests an increase in temperature (with radius) of 14.5 K between 6 and 10 µbar. We also find temperature fluctuations of 1 to 5 K along the path probed by the occultation; if the observed temperature gradients of these fluctuations apply to the vertical direction only, then this region is super-adiabatic. More likely these thermal gradients are due to a combination of diffractive scintillations and horizontal temperature variations. Given that isothermal model fits and numerical inversions cannot separate individual contributions to observed temperature gradients, such as from vertical variations, horizontal variations, and scintillations, this occultation requires a further study.

I. INTRODUCTION

Most of our knowledge of Saturn's vertical temperature structure has come from (i) a series of radio (RSS) occultations, below a radius of 60600 km at a pressure level of 1 bar (Lindal et al. 1985), and (ii) ultraviolet (UVS) occultations above a radius of 61200 km at pressures below the microbar level (Smith et al. 1983) observed with instruments on the Pioneer 11, Voyager 1, and Voyager 2 spacecraft near Saturn. In recent years ground-based stellar occultations have measured Saturn's temperature structure at the microbar level through observations of the 28 Sgr occultation by Saturn in 1989 (Hubbard et al. 1997, hereinafter H97), observations of the central flashes during the 28 Sgr occultation (central flashes probe at much higher pressures—Nicholson et al. 1995), and a series of Hubble Space Telescope (HST) Guide Star Catalog (GSC) occultations by Saturn (Bosh and McDonald 1992) during the years 1994-1996. One of these events is described here, with more information given by Cooray (1997).

Even though Saturn's equatorial region has been well studied by spacecraft instruments, less is known about its north polar latitudes. This region has been observed recently with HST for the purposes of studying haze distributions (Karkoschka and Tomasko 1993), the polar hexagon (Sanchez-Lavega et al. 1996), and auroral activity (Trauger et al. 1994). For Saturn, the existence of two features near the north polar region—the hexagonal wave structure (Allison et al. 1990; Sanchez-Lavega et al. 1993) and the north polar spot (Sanchez-Lavega et al. 1997)—makes it interesting to study the surrounding regions in order to learn about the dynamics associated with them. The atmospheric temperature variations may provide clues to some of the dynamical processes associated with the atmospheric structures.

Rapid rotation with a period of 10.65667 hr (Desch and Kaiser 1981) causes Saturn's atmosphere to be extremely oblate with an oblateness of $0.098 \pm 0.001$ at the 1-bar level (Lindal et al. 1985). Due to the very high zonal winds, the rotation is not uniform and currently the shape is constrained at the 1-bar level by radio occultation observations with various spacecraft (Lindal et al. 1985). Central-flash observations at infrared wavelengths during the 28 Sgr occultation have provided some information on Saturn's atmospheric shape at a pressure level of 250 mbar (Nicholson et al. 1995). The analysis of the total 28 Sgr occultation data set by H97 has shown evidence for the existence of the bulge produced by the high zonal winds at the microbar level. In order to constrain the planetary figure to a higher accuracy, it is necessary that observations be made over a large range of latitudes so that the isopycs (surfaces of constant number density) can be mapped over the entire planet. None of the prior occultation observations, neither visible nor radio, probed Saturn at latitudes higher than 36.3 degrees in the northern hemisphere and 74.2 degrees in the southern hemisphere.

In this paper our goals are to obtain the thermal structure of the north polar region and explore the possible latitudinal temperature variations. The half-light radius and corresponding pressure of this high-latitude region can be used to constrain the shape of Saturn at the microbar level.

## II. OVERVIEW OF THE OCCULTATION

A recent stellar occultation by Saturn's north polar region of the star GSC5249-01240 (V=11.9) probed a long horizontal component of the atmosphere, along with the usual vertical component. This event was considered noteworthy when it was first identified because it had the slowest shadow velocity of all the 203 predicted stellar occultations predicted by Bosh and McDonald (1992). The occultation was predicted to occur during the solar ring-plane crossing of Saturn on November 20-21, 1995, with a relative velocity of only 0.58 km sec$^{-1}$ between the Earth and Saturn in the sky plane (Bosh and McDonald 1992). The advantage of such low velocity occultations is that a high spatial resolution at Saturn can be easily achieved, even with the typically slow array integration times of 1 sec. Since occultations usually occur at velocities of 15 to 20 km sec$^{-1}$, observations of a very low velocity occultation such as this would then allow a signal-to-noise ratio per kilometer in the shadow plane at least 6 times that for a star of the same magnitude, but having a more typical shadow-plane velocity.

Figure 1 shows the calculated path of the primary star (in the rest frame of Saturn) during the occultation, as recreated by a preliminary ring-orbit solution of the subsequent occultation of this star by Saturn's ring system on November 21, 1995 (Bosh and Olkin 1996). Table I lists the adopted coordinates of the star under the assumption that the offset between the relative position of the Saturn's rings and the star is entirely due to the star position, rather than the DE130 ephemeris used to calculate the position of Saturn.

The error in the path shown in Figure 1 is estimated to be less than 4 km. As shown in this figure, immersion occurs in the north polar region. Our observations spanned the region of 82.5 to 85 degrees in latitude and 20 to 30 degrees in longitude, as measured from the prime meridian defined by observations of radio emission rotations by Desch and Kaiser (1981) and Kaiser et al. (1984). During immersion the star traversed 660 km along the limb and dropped vertically 88 km into the atmosphere. Thus our observations are composed of both horizontal and vertical structures of this region at an approximately 7.5 to 1 ratio. This uniquely high ratio of horizontal to vertical motion yields the most extensive continuous record of horizontal structure of a giant planet through stellar occultation observations. Also, these are the first occultation observations to probe the polar region of Saturn.

## III. OBSERVATIONS

We observed the immersion event of the GSC5249-01240 occultation at NASA's IRTF using NSFCAM (Shure et al. 1994) at a central wavelength of 2.280 μm with bandwidth of 0.17 μm (FWHM). The integration time of the observations was 0.904 sec (Table II). The same event was also observed with the HST (Bosh and Olkin 1996), and those data are currently being analyzed. These HST observations probed latitudes near 75 degrees north for immersion and near 15 degrees north for emersion. The emersion of this event was also observed at Palomar observatory by French and Nicholson (private communication), which probed a latitude near 20 degrees north. The emersion data from IRTF is of low signal-to-noise (due to very poor daytime seeing) as the event was observed during daylight at Hawaii.

In order to collect data with adequate signal-to-noise ratio, we used NSFCAM in movie-burst mode with a frame size of 160 by 56 pixels (Figure 2). In movie-burst mode, frames are read out and stored in the temporary storage memory. Once the maximum number of frames allowed by the memory capacity (64 Mb) have been recorded, the data are then written to a hard drive. This memory limitation prohibited us from collecting the data needed to construct the full light curve of this event.

In Figure 3 we show the light curve, where we missed a section of the immersion event because of memory limitations. Due to the extremely small relative velocity of Saturn and the star, we were unable to determine exactly the time of immersion beforehand and we collected data continuously, except at places where we had to stop as the memory capacity of NSFCAM had reached the maximum allowed. As shown in Figure 3, such an occasion occurred during the beginning of immersion. The time lost between the end of first data segment, which includes the top portion of the light curve, to the beginning of second data segment is 288.4 sec. This is approximately the time taken to write the data to the hard drive and for a single stare-frame observation that was used to confirm the telescope tracking and focus.

The data integrations in NSFCAM are started at an accurately known GPS time, but the cycle time while the data are gathered is controlled by an internal PC clock, which is not running synchronously with UT. In order to estimate the deviation between the time based on the PC clock and UT time we have taken two approaches. First we recorded a series of subframes in the same configuration as the occultation observation, using an infrared LED that was triggered with the 1 Hz pulse produced by a GPS receiver. The timing analysis (Smith 1995) showed the cycle time between frames to be 949.61 ± 0.10 ms, of which 45.61 ± 0.10 ms was deadtime used to read and clear the array between integrations. Second, the subsequent ring occultation provided timings of ring events for which a preliminary model has been fitted by Bosh and Olkin (1996). These timings suggest a deadtime of 43 ms and a cycle time of 947 ms. The derivation of these value will be included in a future paper on the ring occultation (Olkin, personal communication). We have opted to use the latter values because we have used the information from the ring timings to recreate the path of the star during the atmospheric occultation. It is not yet clear why the two methods give two different answers for the cycle time. But the difference of 2 msec between the two timings methods produces an error of only 4 km in the half-light radius. As we find later, this is smaller than the error due to model fits.

Two frames are shown in Figure 2. We planned our observations to observe the mutual events by Saturnian satellites simultaneously with the occultation. This forced us to use a large frame size for observations (see Table II), which in turn reduced the number of frames that could be recorded continuously. From the data frames we extracted a subframe around the occulting star of size 8 by 8 pixels (2.4 arcsec by 2.4 arcsec) which was centered such that no contribution from the stellar companion northwest of GSC5249-01240 was introduced. The center of the subframe was set at a constant offset from this companion. The signal recorded in this subframe is the total from occulting star GSC5249-01240 and Saturn's limb. Since the observations were made in a deep $CH_4$ absorption band, most of the Saturn's limb appears dark. We established that the contribution from Saturn's limb is less than 0.1% by examining another subframe of the same size, next to the one that was used for photometry. Also, variations in background flux were monitored by another subframe of the same size. These background counts were subtracted from the signal counts in constructing the light curve shown in Figure 3. As part of our photometry, we also extracted the signal from the companion by including a subframe around it, but we did not use these data to normalize the light curve.

## IV. ANALYSIS

### A. Isothermal Model Fits

In Figure 3 we present the model fit to the light curves based on isothermal model fits as described in Elliot and Young (1992, hereinafter EY92). We refer the reader to EY92 for details regarding isothermal model fits to observed stellar occultation light curves, and derivation of atmospheric parameters based on model fits. The only difference between our analysis and that of

EY92 is that we have carried throughout the rotational contribution to centrifugal force, whereas EY92 ignores it since it was unnecessary for their analysis of Pluto atmospheric data.

We determined the half-light radius based on the $(f,g)$ coordinate system as formulated by Elliot et al. (1993). The astrometric information used is listed in Table I. First, we define the shadow plane, also known as the fg plane (Elliot et al. 1993) to be perpendicular to the line-of-sight to the star, such that the perpendicular to the fg plane, $\hat{h}$ points from the observer to the GSC5249-01240, $\hat{f}$ points in the direction of increasing right ascension and $\hat{g}$ completes the right-handed system (increasing declination). We refer the reader to Elliot et al. (1993) for coordinate transfer equations related to the fg coordinate system, and the derivation of astrometric information based on a ring occultation. The position of the IRTF relative to the center of the Saturn in the fg plane was derived from a preliminary solution of the ring occultation by Bosh and Olkin (1996) and was used to compute the f and g values along the path of the star. With these fg values we express flux values as a function of shadow-plane radius, rather than time, and performed isothermal fits to the normalized signal vs. radius as described in EY92. This fitting procedure was used to determine the full level, the background level, the half-light radius ($\rho_h$), and the ratio ($\lambda_h$) of the absolute value of potential energy to kT (EY92), where k is Boltzmann's constant and T is the equivalent isothermal temperature of the atmosphere.

The parameter $\rho_h$ is the radius of the half-light level in the shadow plane, which can be written as:

$$\rho_h = \left[(f_h - f_0)^2 + (g_h - g_0)^2\right]^{\frac{1}{2}}, \tag{1}$$

where we have evaluated $f(t)$ and $g(t)$, the sky plane coordinates (east and north on the sky respectively), at the time of half-light ($t = t_h$):

$$f_h \equiv f(t_h) \text{ and } g_h \equiv g(t_h). \tag{2}$$

In Eq. (1), $f_0$ and $g_0$ represent the fg component values for the ephemeris correction determined from fit to the ring occultation observations (Elliot et al. 1993). Both of these are zero for our analysis, since the star position used here was corrected, as determined by the ring fit (Bosh and Olkin 1996).

The same calculation can be done by transforming the fg coordinate system in the shadow plane to a coordinate system in the body plane based on the formulation by Elliot et al. (1993). If measured in the two planes, shadow and body, the two half-light radii differ by an amount equal to scale height of the planetary atmosphere ($H_h$) and the deflection due to general relativity $(\Delta r_{grav})$. The first term is due to the refractive bending of the half-light ray from planet plane to the observer plane. The amount due to general relativistic deflection can be written as:

$$\Delta r_{grav} = \frac{4GM_s D}{rc^2}, \tag{3}$$

where $G$ is gravitational constant, $M_s$ is the mass of Saturn ($GM_s = 37931246.375$ km$^3$ s$^{-2}$; W. M. Owen, private communication, as tabulated in Elliot et al. 1993), $c$ is the speed of light, and $D$ is the topocentric distance to the sky-plane at Saturn at the time of observation. For an oblate planet like Saturn, the minimum atmospheric radius probed does not lie in the plane of the sky but in a plane either behind or in front by a small amount. This is a geometric correction that was first included by Baron et al. (1989) in the derivation of the oblateness of Uranus. Using their result and adding the corrections for general relativity, refraction, and oblateness, we can write the half-light radius, r$_h$:

$$r_h = \left(\rho_h + \Delta r_{grav} + H_h\right)\left[1 + \gamma^2 \frac{\sin^2 \phi_h}{\cos^2 B_s}\right], \tag{4}$$

where $\phi_h$ is the planetocentric sub-occultation latitude at the half-light level, which is a function of the position angle of the pole $P_s$ and sub-Earth latitude $B_s$ (Elliot et al. 1993). In Eq. (4) the factor $\gamma$ is given by the equation:

$$\gamma = \frac{-\varepsilon\left(1 - \frac{\varepsilon}{2}\right)\sin 2B_s}{1 - \varepsilon(2 - \varepsilon)\cos^2 B_s}. \tag{5}$$

Here $\varepsilon$ is the oblateness of the planet. We have used the oblateness at the 1-bar level (0.098 ± 0.001; Lindal et al. 1985) derived through spacecraft radio occultations in our analysis of the microbar region of the planet. Eqs. (4) & (5) describe an ellipsoid of revolution, which may not completely describe real Saturn with an extra equatorial bulge due to zonal winds. Our approximated equations should be valid for the present analysis, and is not expect to produce results different by more than 1%.

At the half-light level we can write refractivity as:

$$\nu_h = \frac{H_h^{3/2}}{D\sqrt{2\pi r_{ch}}}, \qquad (6)$$

where we have used the radius of curvature at the half-light level, $r_{ch}$, instead of approximating it as the half-light radius. The radius of curvature here is defined in the plane containing the center of the planet and the line of sight. Based on the geometry, the three-dimensional Saturn can be projected into a two-dimensional ellipse, and the radius of curvature can be evaluated based on the elliptical shape of Saturn. Alternatively one can use the knowledge of location of each of the rays and calculate the phase shift based on the path traveled by the ray (H97). Using the radius of curvature instead of the radius changes the refractivity at the half-light level by about 3 to 4%, depending on the latitude.

The number density, $n_h$, at the half-light level is:

$$n_h = \frac{N_L \nu_h}{\nu_{STP}}, \qquad (7)$$

where $N_L$ is Loschmidt's number and $\nu_{STP}$ is the refractivity at standard temperature and pressure. For Saturn's atmosphere at the 1-μbar level we have adopted a hydrogen abundance of 94% based on the Conrath et al. (1984) interpretation of infrared data obtained with spacecraft. The remainder is assumed to be helium, with a small amount of methane. The mean molecular mass of this mixture is 2.135 amu with an uncertainty of about 2.5%. At the wavelength of our observations this gas mixture corresponds to a $\nu_{STP}$ of about 1.36×10⁻⁴ (Elliot et al. 1974; Peck and Huang

1977). Using the number density at the half-light level, we can write the pressure at the half-level as:

$$p_h = n_h k T_h, \qquad (8)$$

where $k$ is the Boltzmann's constant and $T_h$ is the isothermal temperature of the planet which is related to the scale height, $H_h$, and local acceleration of gravity, $g$, through:

$$T_h = \frac{H_h \mu m_{amu} g}{k}. \qquad (9)$$

Using the results of our model fits and making the above described corrections for general relativity, refractive bending, and geometric effects—we find the half-light intensity level to occur at a radius ($r_h$) of 54964 ± 10 km. The corresponding energy ratio for the isothermal fit ($\lambda_h$) is 1254 ± 120 (Table III). The scale height at this level is 42.8 ± 4.3 km, corresponding to an isothermal temperature of 130 ± 10 K, based on gravitational acceleration value of 1187 cm sec$^{-2}$. The gravitational acceleration was calculated from the rotation and zonal winds as described in Appendix A of Lindal et al. (1985). This radius, at the half-light intensity level, corresponds to a pressure level of 1.61 ± 0.10 μbar, with a number density of $(8.80 \pm 0.07) \times 10^{13}$ cm$^{-3}$.

Based on isothermal fits to the 28 Sgr occultation observations in 1989, H97 found a mean equatorial temperature of 141 ± 4 K, with a scale height of 62.3 ± 4.8 km. Note that the scale heights differ much more than the temperatures, due to the smaller half-light radius and lower centrifugal acceleration at the pole compared with the equator. Within the errors, the temperature we found for GSC5249-01240 light curve at the north polar region agrees with the equatorial temperature. Also our derived values for pressure and number density are in perfect agreement with "L+Inv" model of H97. Assuming that the computed values are on a level surface with an equivalent equatorial radius (see H97 for details), one get a value of 60987 km with corresponding pressure of 1.62 μbar, temperature of 134 K, and number density of $8.80 \times 10^{13}$ cm$^{-3}$, which agrees with the derived the values from isothermal models in this paper. The agreement of H97's global temperature model, derived from the equatorial data, with our observations of the north polar region argues against large latitudinal temperature variations in Saturn.

B.  Numerical Inversions

We applied the inversion algorithm described by Wasserman and Veverka (1973) and French et al. (1978), with modifications by Elliot (unpublished notes, 1989), Olkin (1996) and Cooray (1997) to invert the observed light curve to derive the thermal profiles.  The only difference between published work and our modified procedure is that we have avoided the large-planet approximations.  Specifically, we do not assume that the scale height is much less than the radius and we include horizontal focusing by limb curvature.  This form was developed for inversion of small-planet occultation curves (Olkin 1996), and in this case the results should be virtually indistinguishable from the published methods.  In order to test the sensitivity of our numerical inversion procedure, we performed the inversion with various initial conditions.  Since we were missing a section of the light curve, the inversion was initialized at levels very close to the level at which we have data.  Before the inversion, the light curve was renormalized based on the values of full level and background level from the isothermal fit.  Changing the background values within 10% of the value derived did not vary the temperature profiles significantly. Only a large change (25%) in the background altered the observed profile, decreasing the observed temperature gradient and producing a temperature profile with a near-isothermal atmosphere.  Also, changing the background did not change the observed super-adiabatic features discussed below, which are primarily due to the high amplitude spikes in the occultation light curve.  To initiate the inversion, we performed isothermal fits to the renormalized light curve and established radius values and the refraction angle at various flux levels, from 0.48 to 0.42.  These different values were considered as possible starting places for inversion, with the calculated radii and refraction angles for each of these flux values as initial conditions. The derived temperature profiles for different starting points are shown in Figure 4.  As shown all the temperature profiles converge to a temperature value of 114.6 ± 2.4 K at a pressure level of 13.5 ± 0.1 μbar.  We have shown typical error bars throughout these profiles, which were calculated based on the formulation by French et al. (1978), slightly modified to our inversion procedure (Cooray 1997).

In comparison with the isothermal model fits, temperature profiles from the inversion suggest temperatures that agree at the top level of 1.5 to 2 µbar. As we probe deeper into the atmosphere, a main feature on the temperature profile is a warm layer between the 6- and 10-µbar level, where the temperature decreases by 14 K within a vertical distance of 18 km, and over a horizontal distance of 170 km. If the entire change occurred in the vertical direction, the temperature gradient would be $0.8 \pm 0.1$ K km$^{-1}$. On the other hand, if the entire gradient were horizontal, then the average gradient would be $0.09 \pm 0.01$ K km$^{-1}$. Also seen are temperature variations on small vertical scales, corresponding to the high intensity spikes in the occultation light curve. We will investigate reasons for the high temperature gradient in Section V. B. and reasons for the small scale variations in Section V. C.

## V Discussion

A. Seasonal and Latitudinal Temperature Variations

In Figure 5 we have shown all the temperature measurements obtained so far with stellar occultations by Saturn's atmosphere. These values include the temperatures obtained for the 1989 28-Sgr occultation by H97 and observations of the GSC5249-01240 immersion. The temperatures determined by Cooray (1997) and H97 for the 28 Sgr immersion observations from IRTF agree within 1%, confirming the consistency of two different methods of analysis.

A theoretical model based on a multilayer radiative transfer model applied to data obtained with spacecraft illustrates the expected seasonal cycle of the stratospheric north polar, south polar and equatorial temperatures at pressure level of 5 mbar (Bezard and Gautier 1985, esp. their Fig. 5). Their predicted maximum north-pole to equator temperature difference is 20 K, which was calculated to have occurred just prior to the Voyager encounters. In 1995, the north polar region was calculated to be warmer by at least 8 K with respect to the equatorial temperature, whereas the equatorial temperature did not vary more than 2 K throughout the time since 1981. If the region from 5 mbar to 1 µbar is isothermal, this predicts a polar temperature of 149 K, which is greatre than our measurement by two sigma.

When temperature profiles from inversions are considered, our observations suggest that the north polar region is 15 K colder at pressure levels below 10 µbar when compared with temperatures at a pressure of 1.5 µbar, suggesting that the north polar region between 5 mbar and 1 µbar is in fact not isothermal. This is contrary to what is assumed in H97, where they have interpolated temperature values at 5 mbar obtained from radio occultations to values at 1 µbar from stellar occultations, assuming an isothermal atmosphere. It is still uncertain if Saturn's global temperature structure is isothermal from the millibar region (where current theoretical models exist) to the microbar region (where stellar occultations are sensitive).

B. Thermal-Gradient Feature

An interesting feature in the temperature profile is the large thermal gradient feature observed between the vertical distances of 54840 and 54860 km (less than a scale height) where the temperature increases with altitude by 14 K. Similar temperature gradients may have been observed in temperature profiles for Uranus and Neptune obtained with visible and infrared occultations (French et al. 1983; Roques et al. 1994). Since these other observations probed mostly a vertical component of the atmosphere, one might conclude that this gradient is also due to vertical variations and not horizontal variations.

A gradient such as this requires a heating mechanism to sustain it. For Neptune, dynamical mechanisms (such as propagating waves) were suggested to explain the rapidly changing aspect of the thermal profiles both in time and amplitude. This large gradient can also be a signature of solar radiation absorption by aerosols. Our inversion procedure assumes a transparent atmosphere, and if there is such absorption, the inversion interprets it as a decrease in temperature producing the observed temperature gradient. However, it is difficult to discriminate solely from inversion between absorption by aerosols and a large temperature gradient of temperature. It is also possible that the aerosols could be dense enough to be important for heating in the atmosphere, but yet thin enough so that the small optical depth wouldn't affect the stellar occultation light curve.

Even though there may be dynamical difficulties in maintaining aerosol layers that may absorb solar radiation and heat the stratosphere, Karkoschka and Tomasko (1993) have observed the higher latitudes of Saturn in search of hazes with HST. Optical depths in the ultraviolet of stratospheric hazes at latitudes above 70° N were found to be almost unity at altitudes extending above the 10-mbar level, whereas no hazes were found in low and mid-southern latitudes at this pressure level. The tropospheric haze distribution is completely different from the distribution of stratospheric haze, since optical depths in the troposphere strongly decreases from the north pole to equator. In addition, Karkoschka and Tomasko (1993) have shown that the stratospheric and tropospheric particles have different sources and compositions.

Given the lack of aerosols in the stratosphere at mid-latitudes, it seems possible that the mechanism of formation of stratospheric particles is different at polar and equatorial latitudes. In the polar region, a mechanism by which the particles are produced at very high latitudes is required. For Uranus, Rizk and Hunten (1990) have shown that a dust mass influx of $10^{11}$ g yr$^{-1}$ spiraling inward from the Uranian ring system can raise the temperatures up to 50 K in the 1-10 µbar region. For Saturn, a magnetic connection between Saturn's rings and atmosphere has been suggested by Connerney (1986). The latitudinal variations in images of Saturn's disk, upper atmospheric temperatures from Voyager observations, and ionospheric electron densities are found in magnetic conjugacy with features in Saturn's ring plane. These variations were described as the result of a variable influx of water, in the form of high-charge-to-mass-ratio ions or submicron grains with one electron charge, transported along magnetic field lines from sources in the ring plane. Feuchtgruber et al. (1997) recently reported a detection of $H_2O$ and $CO_2$ in the upper atmospheres of giant planets, which further suggests the existence of an external supply of water to giant planet atmospheres.

The high zonal winds of Saturn may also be involved in the transportation of aerosols. Karkoschka and Tomasko (1993) have hypothesized that the equatorial stratospheric aerosols are a transient phenomena and that they may relate to the great storms observed near the equatorial regions. A similar process could well exist at higher latitudes, where transportation mechanisms may transfer aerosols from cloud levels found in the millibar to microbar levels in Saturn's atmosphere. Even if there is a vertical transporting mechanism for aerosols, a very efficient confinement method is required to maintain the aerosol layers, and that would need to be investigated for Saturn.

C. Static Stability and Atmospheric Waves

In general an atmosphere is stable to convection if the lapse rate (-dT/dz) is less than the adiabatic lapse rate $g/c_p$, where $c_p$ is the specific heat at constant pressure. As shown in Figure 6, the GSC5249-01240 occultation indicates lapse rates which, if interpreted as vertical, are very close to and sometimes greater than the local adiabatic rate. The temperature gradients derived from the IRTF 28-Sgr immersion temperature profile never reach the adiabatic value (upper curve). The observed super-adiabatic temperature gradients of the GSC5249-01240 profile are correlated with the spikes in the light curve with normalized intensity greater than 1. According to Elliot and Veverka (1976), spikes greater than 1 suggest ray crossing. But the effect of ray crossing will be to smooth out the features in the temperature profile, rather than to produce any spurious variations. None of the prior occultations of giant planets probed at as a high resolution as ours, which is primarily achieved due to small stellar diameter (~0.2 km) projected at the distance to Saturn's atmosphere, much larger than the Fresnel diffraction scale of ~1.8 km. The cycle time of the data recording (0.947 sec, Table 1) corresponds to an averaging distance of 0.55 km in the shadow. However, the corresponding dimension in the atmosphere decreases proportional to the normalized light-curve flux as the occultation proceeds. Hence the limit to the spatial resolution would be Fresnel diffraction.

It is possible that apparent super-adiabatic lapse rates such as the ones observed here have been filtered out in prior occultations because of lower spatial resolution. Such filtering can occur if the observed super-adiabatic gradients are an artifact due to diffractive scintillations. Given the high resolution of our observations, it is possible that we are at a regime where both diffractive and refractive scintillations are important. Prior occultations, such as Neptune occultations analyzed in Narayan & Hubbard (1988), were limited by the stellar diameter. In such occultation light curves, authors found the presence of refractive scintillations. The existence of both refractive and diffractive scintillations can be tested by averaging out the light curve.

We have tested the dependence of the super-adiabatic gradients on the time resolution of our data by constructing an average light curve, in which each point represents a five-point average of the original light curve. This produces the light curve that would have been observed if the cycle time of one frame was five times longer than the value we used. As shown in Fig. 7 such averaging takes out the rapid variations in the temperature gradients, and indicates a stable atmosphere. Such temperature variations are similar to those found for prior occultations, which were in the refractive regime.

If the observed temperature gradients are interpreted as vertical gradients and real, they imply a super-adiabatic state; therefore it is highly likely that most of these temperature gradients are either horizontal and not vertical, or an effect due to diffractive scintillations. If these are in fact real and vertical gradients, then the north polar atmosphere is in a state not seen elsewhere. Super-adiabaticity is ordinarily quickly erased by turbulent convection. Such super-adiabatic gradients have not been observed in any of the temperature profiles derived for giant planets based on stellar occultations so far, but none of these occultations probed as long a horizontal structure as we did. Currently no theory exists that would allow us to clearly state how much of the variations are horizontal, but we can set an upper limit on the horizontal variations by ignoring the contributions due to scintillations and vertical variations. In Fig. 8 we show the temperature gradients under the assumption that all the temperature variations are due to horizontal structure. Horizontal variations could be produced by a combination of gravity waves and vertical wind shear. However, the gradients in Fig. 8 are not true horizontal temperature gradients because the inversion technique assumes only vertical variations in atmospheric structure, and new techniques would be needed to cope with horizontal variations in atmospheric structure. One could set a lower limit on the variations due to horizontal structure by subtracting putative vertical temperature gradients that do not exceed the adiabatic value. However, in this case such a calculation should also take into account the effects due to diffractive scintillations, which can produce rapid variations in the temperature profile.

At this stage, we are unable to state the reasons for the observed temperature variations, either horizontal or vertical, of the profile, and whether they are due to scintillations. It is certainly possible that most of the observed high-frequency variations are due scintillations.

## VI CONCLUSIONS

From the GSC5249-01240 occultation observations we find the equivalent isothermal temperature of Saturn at a half-light latitude of 83°.2 N and pressure of $1.61 \pm 0.09$ µbar is $130 \pm 10$ K. The radius of Saturn at this half-light level is $54964 \pm 10$ km. This half-light radius gives information to constrain the planetary figure and study the effect of high zonal winds on Saturn at the pressure level of microbars. We leave this task to be carried out with the observed immersion events from HST of the same occultation. The temperature of the north-polar region is comparable to the equivalent isothermal temperature near the equator, and this can aid the global thermal modeling of Saturn's atmosphere. The temperature profile based on numerical inversion suggests an increase in temperature of 14 K over a vertical length less than a scale height between 6 and 10 µbar, which requires a heating mechanism. We have considered the possibility of the presence of aerosols in the stratosphere, as the source of energy. The observed spikes in the GSC5249-01240 light curve should be modeled based on the gravity waves in this region and scintillation theory.

We may have detected horizontal temperature variations of the north polar region, where Saturn's polar hexagon has become an interesting topic over the last few years for both observational and theoretical studies (Sanchez-Lavega et al. 1993). As planetary-occultation studies in the past were directly involved in the study of vertical atmospheric profiles, and now we have observed a nearly horizontal component, new methods need to be investigated that will allow us to separate the vertical component from the horizontal component and study temperature variations separately. Also given the resolution of this occultation, where we might be probing the atmosphere in such a manner that diffractive scintillations are important, this occultation requires a further study based on scintillation theory. As discussed, due to instrument memory limitations we have lost a significant amount of data from the light curve. We hope that in the near future instruments at major observatories will have better capabilities for recording long, uninterrupted time series for occultation observations.

## VII ACKNOWLEDGMENTS

We thank the IRTF support staff and telescope operators for help in our observations. We also acknowledge the help received from Cathy Olkin in model fitting and inversion procedures., and our reviewers Bill Hubbard and Francoise Roques for detailed comments on our manuscript. Partial support for this work was provided by NASA through grant number GO-05824.01-94A (Lowell Observatory) from the Space Telescope Science Institute, which is operated by AURA, Inc., under NASA contract NAS 5-26555, and partial support was also provided by NSF Grant No. AST-9322115 (MIT).

References

Allison, M., D. A. Godfrey, and R. F. Beebe 1990. A wave dynamical interpretation of Saturn's polar hexagon. Science 247, 1061-1063.

Baron, R. L., R. G. French, and J. L. Elliot 1989. The oblateness of Uranus at the 1-mbar level. Icarus 78, 119-130.

Bezard, B., and D. Gautier 1985. A seasonal climate model of the atmospheres of the giant planets at the Voyager encounter time. Icarus 61, 296-310.

Bosh, A. S., and S. W. McDonald 1992. Stellar occultation candidates from the Guide Star Catalog. I. Saturn, 1991-1999. Astron. J. 103, 983-990.

Bosh, A. S., and C. B. Olkin 1996. Low optical depth features in Saturn's rings: The occultation of GSC5249-01240 by Saturn and its rings. Bull. Amer. Astron. Soc. 28, 1124.

Connerney, J. E. P. 1986. Magnetic connection for Saturn's rings and atmosphere. Geophys. Res. Lett. 13, 773-776.

Conrath, B. J., D. Gautier, R. A. Hanel, and J. S. Hornstein 1984. The Helium abundance of Saturn from Voyager measurements. Astrophys. J. 282, 807-815.

Cooray, A. 1997. Stellar Occultation Observations of Saturn's Upper Atmosphere. M. S. Thesis, Department of Earth, Atmospheric, and Planetary Sciences, Massachusetts Institute of Technology.

Desch, M. D., and M. L. Kaiser 1981. Voyager measurement of the rotation period of Saturn's magnetic field. Geophysics Research Letters 8, 253-256.

Elliot, J. L., A. S. Bosh, M. L. Cooke, R. C. Bless, M. J. Nelson, J. W. Percival, M. J. Taylor, J. F. Dolan, E. L. Robinson, and G. W. van Citters 1993. An occultation by Saturn's rings on 1991 October 2-3 observed with the Hubble Space Telescope. Astron. J. 106, 2544-2572.

Elliot, J. L., and J. Veverka 1976. Stellar occultation spikes as probes of atmospheric structure and composition. Icarus 27, 359-386.

Elliot, J. L., L. H. Wasserman, J. Veverka, C. Sagan, and W. Liller 1974. The occultation of Beta Scorpii by Jupiter. II. The hydrogen-helium abundance in the Jovian atmosphere. Astrophys. J. 190, 719-729.

Elliot, J. L., and L. A. Young 1992. Analysis of stellar occultation data for planetary atmospheres. I. Model fitting, with application to Pluto. Astron. J. 103, 991-1015.

Feuchtgruber, H., E. Lellouch, T. de Graauw, B. Bezard, T. Encrenaz and M. Griffin 1997. External supply of oxygen to the atmospheres of giant planets. Nature. 389, 159-162.

French, R. G., J. L. Elliot, E. W. Dunham, D. A. Allen, J. H. Elias, J. A. Frogel, and W. Liller 1983. The thermal structure and energy balance of the Uranian upper atmosphere. Icarus 53, 399-414.

French, R. G., J. L. Elliot, and P. J. Gierasch 1978. Analysis of stellar occultation data. Effects of photon noise and initial conditions. Icarus 33, 186-202.

Hubbard, W. B., C. C. Porco, D. M. Hunten, G. H. Rieke, M. J. Rieke, D. W. McCarthy, V. Hammerle, J. Haller, B. McLeod, L. A. Lebofsky, R. Marcialis, J. B. Holberg, R. Landau, L. Carrasco, J. Elias, M. W. Buie, E. W. Dunham, S. E. Persson, T. Botoson, S. West, R. G. French, J. Harrington, J. L. Elliot, W. J. Forrest, J. L. Pipher, R. J. Stover, B. Sicardy, and A. Brahic 1997. Structure of Saturn's mesopshere from the 28 Sgr occultations. submitted.

Kaiser, M. L., M. D. Desch, W. S. Kurth, A. Lecacheux, F. Genova, B. M. Pedersen, and D. R. Evans 1984. Saturn as a radio source. In Saturn (T. Gehrels and M. S. Matthews, Ed.), pp. 378-415. University of Arizona Press, Tucson.

Karkoschka, E., and M. G. Tomasko 1993. Saturn's upper atmospheric hazes observed by the Hubble Space Telescope. Icarus 106, 428-441.

Lindal, G. F., D. N. Sweetnam, and V. R. Eshleman 1985. The atmosphere of Saturn: an analysis of the Voyager radio occultation measurements. Astron. J. 90, 1136-1146.

Narayan, R., and W. B. Hubbard 1988. Theory of anisotropic refractive scintillation: Application to stellar occultations by Neptune. Astrophys. J. 325, 503-518.

Nicholson, P. D., C. McGhee, and R. G. French 1995. Saturn's central flash from the 3 July 1989 occultation of 28 Sgr. Icarus 113, 57-83.

Olkin, C. B. 1996. Stellar occultation studies of Triton's atmosphere. Ph. D. Thesis, Massachusetts Institute of Technology.

Peck, E. R., and S. Huang 1977. Refractivity and dispersion of hydrogen in the visible and near-infrared. J. Opt. Soc. Am. 67, 1550-1554.

Rizk, B., and D. M. Hunten 1990. Solar heating of the Uranian mesopause by dust of ring origin. Icarus 88, 429-447.

Roques, F., B. Sicardy, R. G. French, W. B. Hubbard, A. Barucci, P. Bouchet, A. Brahic, J.-A. Gehrels, T. Gehrels, I. Grenier, T. Lebertre, J. Lecacheux, J. P. Maillard, R. A. McLaren, C.

Perrier, F. Vilas, and M. D. Waterworth 1994. Neptune's upper stratosphere, 1983-1990: Ground-based stellar occultation observations. Astron. & Astrophys. 288, 985-1011.

Sanchez-Lavega, A., J. Lecacheux, F. Colas, and P. Laques 1993. Ground-based observations of Saturn's north polar spot and hexagon. Science 260, 329-332.

Sanchez-Lavega, A., J. F. Rojas, J. Lecacheux, F. Colas and P. V. Sada 1997. New observations and studies of Saturn's long-lived north polar spot. Icarus. 128, 322-334.

Shure, M., D. W. Toomey, J. T. Rayner, P. M. Onaka, and A. T. Denault 1994. NSFCAM: A new infrared array camera for the NASA Infrared Telescope Facility. Instrumentation in Astronomy VIII, Kona, HI, pp 25-33.

Smith, G. R., D. E. Shemansky, J. B. Holberg, A. L. Broadfoot, B. R. Sandel, and J. C. McConnell 1983. Saturn's upper atmosphere from the Voyager 2 EUV solar and stellar occultations. J. Geophys. Res. 88, 8667-8678.

Smith, J. D. 1995. Lunar Occultation Timing in a Two-Telescope Observation of Saturn. S.B. Thesis, Massachusetts Institute of Technology.

Wasserman, L. H., and J. Veverka 1973. On the reduction of occultation light curves. Icarus 20, 322-345.

Table I  Occultation Astrometric Parameters

| Parameter | Value |
| --- | --- |
| $a_s$ (J2000, FK5) | $23^h\ 19^m\ 34^s.5947$ |
| $d_s$ (J2000, FK5) | $-6°\ 47'\ 10''.4411$ |
| $\langle v_\perp \rangle$ (km sec$^{-1}$) | 0.586 |
| $f_h - f_0$ (km) | $-910.42$ |
| $g_h - g_0$ (km) | 54901.96 |
| $P_s$ (degrees) | 5.02 |
| $B_s$ (degrees) | 2.67 |
| D (AU) | 9.194 |
| $\phi_h$ (degrees) | 83.2±0.1 |

Table II  Instrument and Observational Parameters

| Parameter | Value |
|---|---|
| Camera | NSFCAM |
| Stored Frame Size (pixels) | 160×56 |
| Pixel Scale (arcsec pixel$^{-1}$) | 0.306 |
| Integration Time (sec) | 0.904 |
| Cycle Time (sec) | 0.947 |
| Central Wavelength (µm) | 2.280 |
| Filter Bandwidth FWHM (µm) | 0.17 |

Table III  Parameters Derived from Isothermal Model Fits

| Parameter | Value |
|---|---|
| Background Level | –0.044±0.004 |
| Full Level | 1.000±0.004 |
| $\rho_h$ (km) | 54913.9±8.5 |
| $H_h$ (km) | 42.8±4.4 |
| $\Delta r_{grav}$ (km) | 41.4 |
| $\gamma$ | –0.011 |
| $r_h$ (km) | 54964.4±9.7 |
| $\lambda_h$ | 1254±120 |
| $g$ (cm sec$^{-2}$) | 1187 |
| $T_h$ (K) | 130±10 |
| $p_h$ (µbar) | 1.61±0.09 |
| $n_h$ (cm$^{-3}$) | $(8.80\pm0.07)\times10^{13}$ |

FIGURE CAPTIONS

Figure 1. Path of GSC5249-01240 through Saturn's atmosphere and ring system on 1995 November 20-21. The geometry of this event was computed based on the times of circular ring events observed during the subsequent ring occultation and the Saturn's pole position (Bosh and Olkin 1996). The uncertainty in this construction is less than 4 km. The immersion was observed in the north polar region of Saturn from 82.5 to 85 degrees in planetocentric latitudes and 10 to 20 degrees in longitudes. The thick part of the trajectory near the north pole corresponds to the time interval of the light curve in Fig. 3.

Figure 2. Data frame of the Saturn system on Nov. 20 UT before (top left panel) and during the Saturn immersion occultation of GSC5249-01240 (bottom left panel). The data were recorded using the NSFCAM with 2.280 μm filter. North is down, east is to the right. Several Saturnian satellites are visible on the right side of the frame. The two stars, GSC5249-01240 and its companion, are on the northern hemisphere of Saturn and the bright star is occulted. The frames on the right show a magnified view of the pixels containing these two stars. Note that the entire northern hemisphere is invisible due to $CH_4$ absorption at the observed wavelength.

Figure 3. The normalized stellar intensity recorded during the immersion occultation of GSC5249-01240 by Saturn at NASA's IRTF with the NSFCAM. The abscissa are (i) seconds past 05:28 UT on November 20, 1995 and (ii) Saturn's planetocentric latitude in degrees of the occulted star as determined from Saturn's pole position and the sub-Earth latitude. The intensity counts were normalized based on the mean value before and after the event. In our isothermal procedure we allow the full level and background level to vary along with half-light radius and energy ratio.

Figure 4. The inversion algorithm (Elliot 1989; Olkin 1996; Cooray 1997) applied to the November 20, 1995 GSC5249-01240 immersion data. The different temperature profiles correspond to the different initial conditions by varying the starting point of numerical inversion based on different model fits. Even though there is a slight divergence at the top level of the inversion, the values converge at the bottom to a steady temperature of 115 K. The thicker line in the middle is the preferred temperature profile, where many profiles converge with different initial conditions. The horizontal lines shown here are the inversion temperature errors, which are primarily due to the error in initial conditions and flux errors, as formulated by French et al. (1978) and Cooray (1997).

Figure 5. Equivalent isothermal temperature at the half-light radius vs. the latitude of the sub-occultation point. We have also shown the values obtained by H97 based on 28 Sgr observations. We find no strong variation in temperature obtained with isothermal fits vs. latitude in Saturn at pressure levels close to 1 to 2 µbar.

Figure 6. Temperature gradients based on the temperature profile of the GSC5249-01240 and the 28 Sgr occultations. Here we have plotted the derivative of the temperature in the altitude range probed and the adiabatic lapse rate as vertical lines. (a) 28 Sgr. The temperature derivatives are never near the adiabatic lapse rate, but are centered around zero suggesting a nearly isothermal atmosphere. (b) GSC5249-01240. For short times scales and altitude ranges less than 3 km, the lapse rate exceeds the adiabatic lapse rate. On average, the gradients are close to the adiabatic lapse rate, suggesting that there are saturated gravity waves in this region. Turbulent activity may be expected at regions where the lapse rate exceeds the adiabatic value; these regions coincide with the observed spikes in the light curve. Note that the vertical range plotted for the 28 Sgr occultation lapse rates is twice as that of the GSC5249-01240 lapse rates. This is due to the fact that our data for GSC5249-01240 occultation began after the half-light level. The pressure level corresponding to the 0 vertical height of the two curves are 18.6 and 14.9 µbars for the 28 Sgr (a) and GSC5249-01240 (b) respectively.

Figure 7. Vertical temperature gradients calculated by replacing five points in the light curve by their mean value. The temperature gradients from the profile show no super-adiabatic values.

Figure 8. Temperature gradients derived from the GSC5249-01240 occultation data under the assumption that all variations are due to horizontal structure. The horizontal gradients never exceed $\pm 1$ K km$^{-1}$, and some of the gradients may be due to vertical variations. However, most of the rapid variations shown here are likely to be due to diffractive scintillations, which needs to be further investigated.

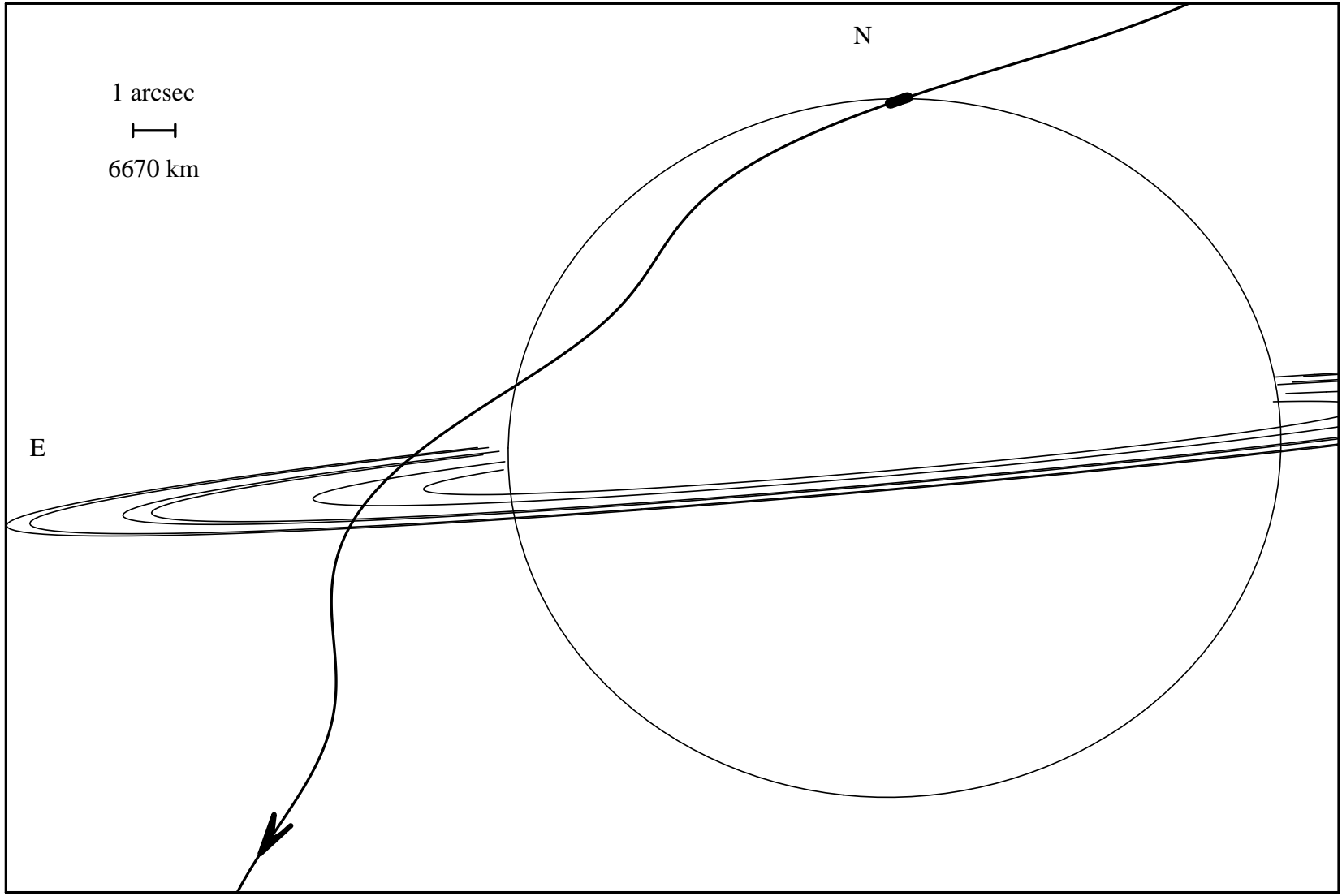

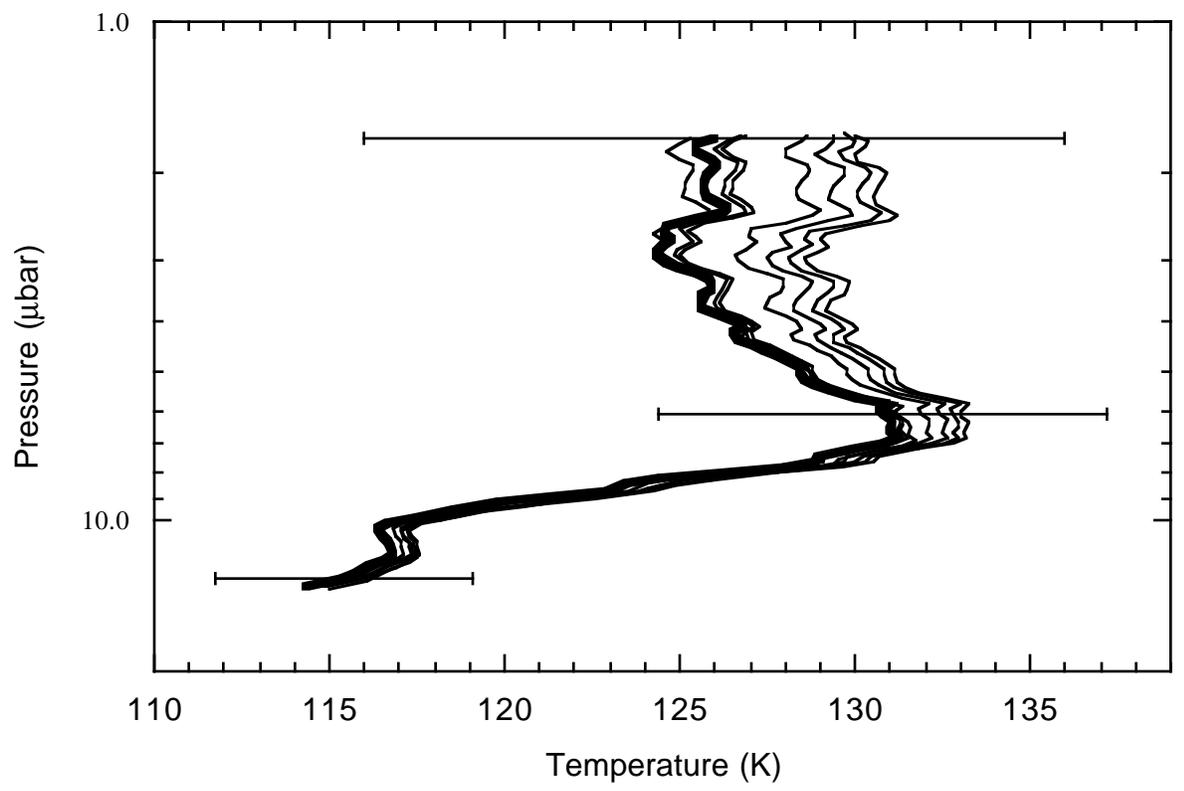

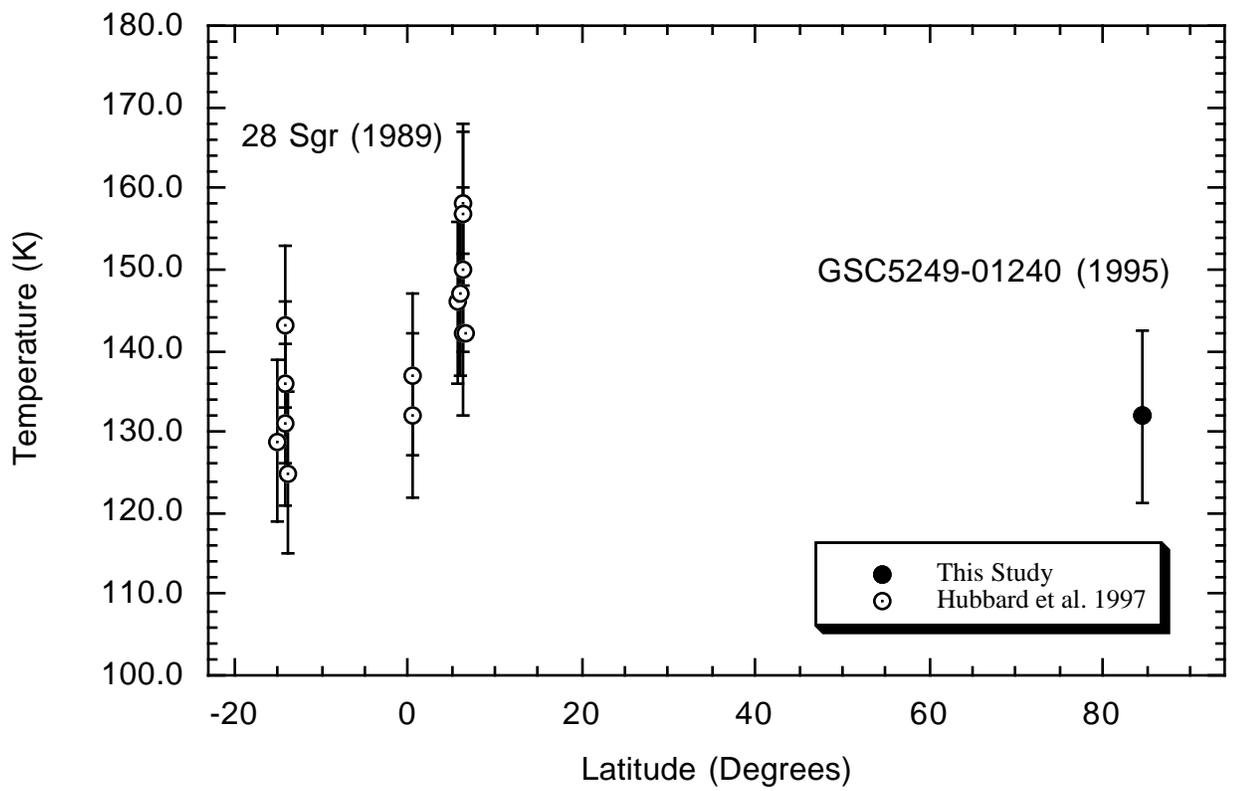

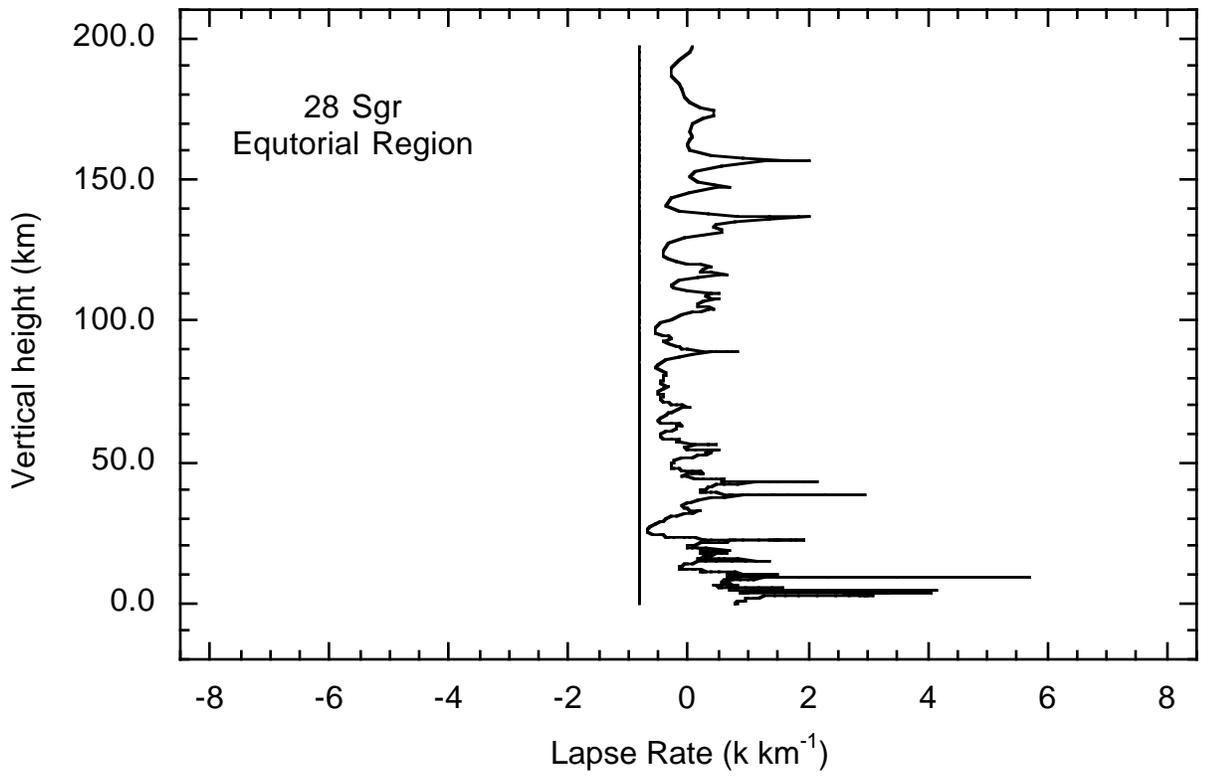

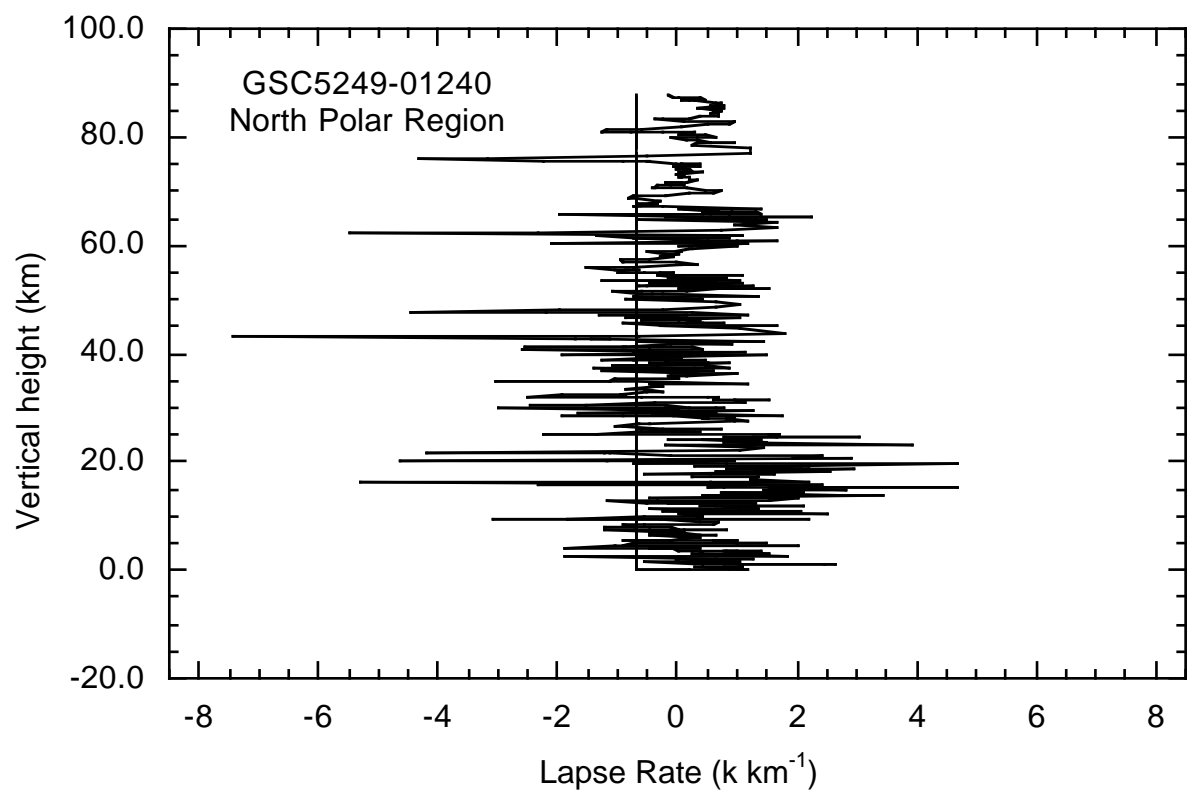

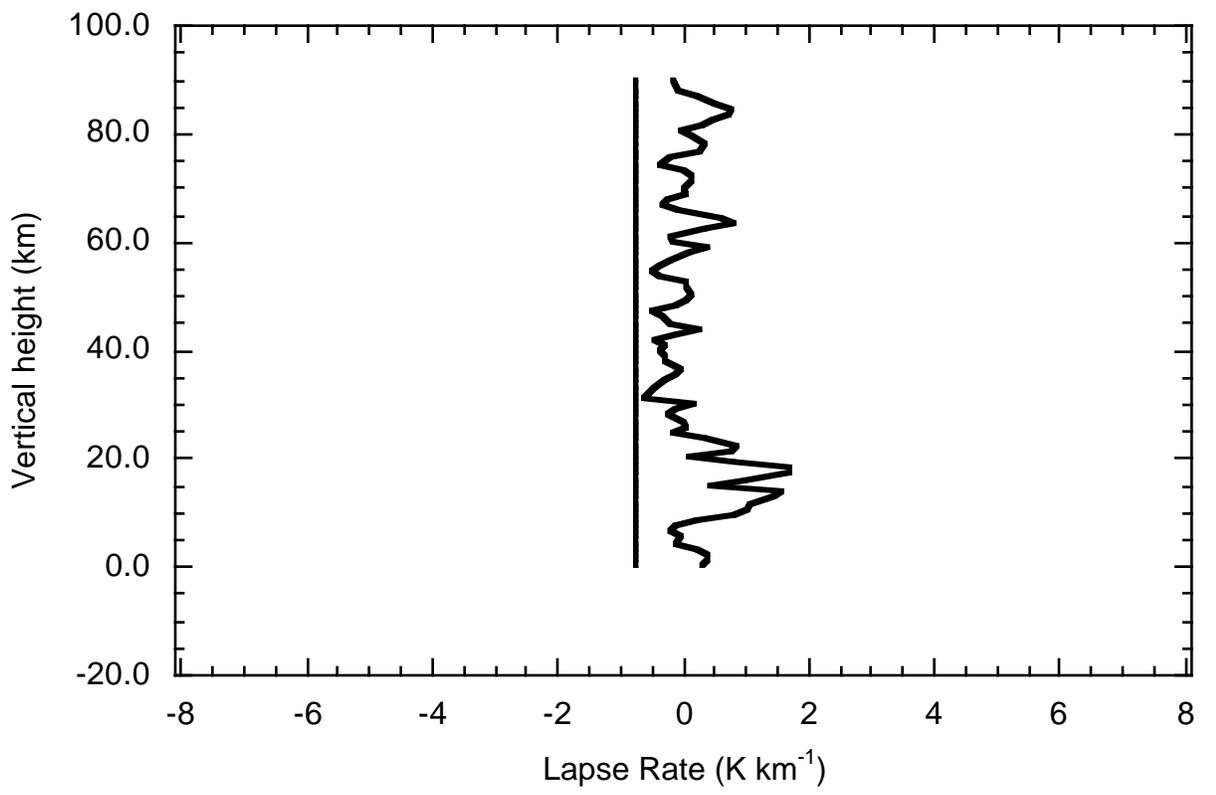

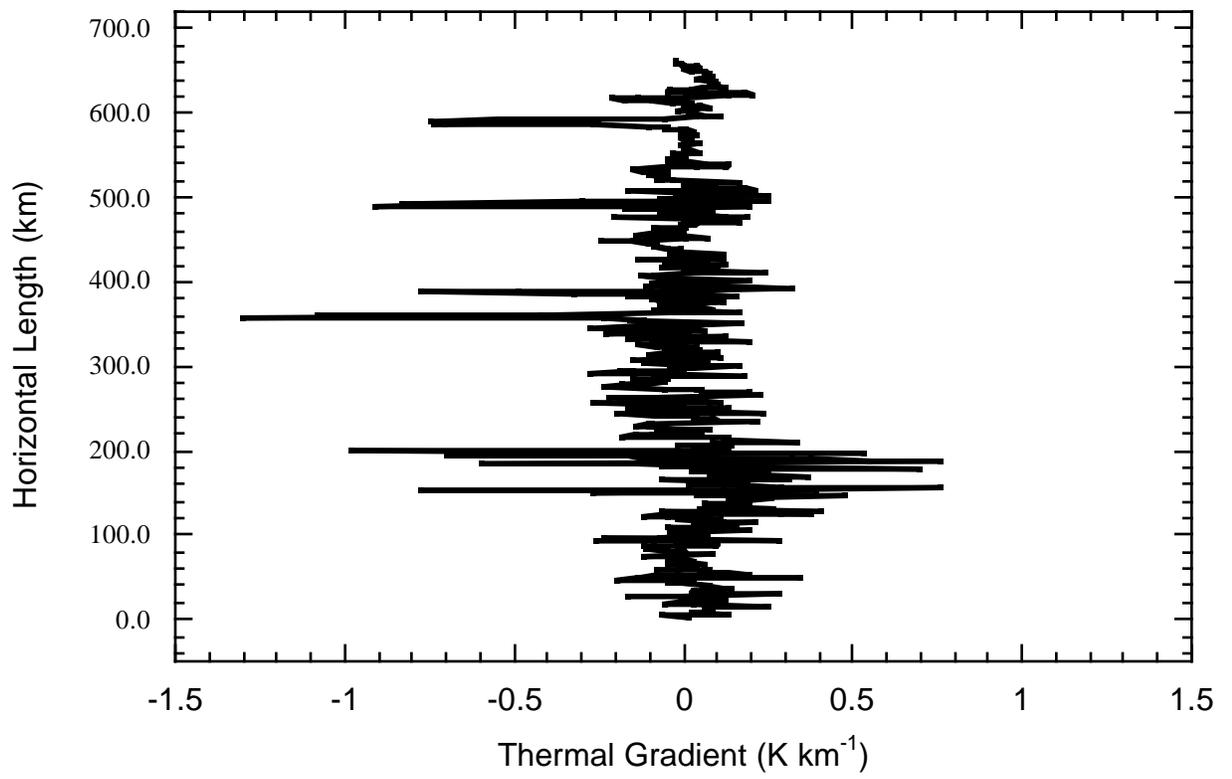